\documentstyle[12pt]{article}
\author{G.N.Parfionov, R.R.Zapatrin\thanks{
Friedmann Laboratory for theoretical physics,
SPb UEF, Griboyedova, 30-32,
191023, St.Petersburg, Russia
}} 

\date{}

\title{Connes duality in Lorentzian geometry}

\def\be{\begin{equation}\label}
\def\ee{\end{equation}}

\newcommand{\1}{{\bf 1}}

\newcommand{\aaa}{{\cal A}}
\newcommand{\ccc}{{\bf C}}
\newcommand{\cinf}{\mbox{$ {\cal C}^{\infty} $}}

\newcommand{\fff}{{\cal F}}

\newcommand{\hhh}{{\cal H}}

\newcommand{\lll}{{\cal L}}

\newcommand{\mmm}{{\cal M}}

\newcommand{\rrr}{{\bf R}}
\newcommand{\vvv}{{\cal V}}

\newcommand{\eprf}{\hspace*{\fill}$\Box$}
\newcommand{\proof}{\paragraph{Proof.}}

\begin{document}

\maketitle

\begin{abstract}
The Connes formula giving the dual description for the distance
between points of a Riemannian manifold is extended to the
Lorentzian case. It resulted that its validity essentially depends
on the global structure of spacetime. The duality principle
classifying spacetimes is introduced. The algebraic account of the
theory is suggested as a framework for quantization along the
lines proposed by Connes. The physical interpretation of the
obtained results is discussed.
\end{abstract}

\section*{Introduction}

The mathematical account of general relativity is based on the
Lorentzian geometry being a model of spacetime. As any model, it
needs identification for a physicist in terms of measurable values.
In this paper we focus on evaluations of intervals between events,
and the measurable entities will be the values of scalar fields.

\medskip

This work was anticipated by the Connes distance formula for
Riemannian manifolds

\[
{\rm Dist}\, (x,y) = \sup\{ f(y) - f(x) \}
\]

\noindent where the supremum is taken over the smooth functions
whose gradient does not exceed 1.

This formula gives rise to a new paradigm in the account of
differential geometry being more sound from the physicist's point
of view since it expresses the distance through the values of
scalar fields on the manifold. Our goal was to investigate to what
extent this formula is applicable in Lorentzian manifolds.

The first observation was that even in the Minkowskian space this
formula is no longer valid in its literal form. The reason is that
the Cauchy inequality on which the Connes formula is based, does
not hold in the Minkowskian space. In section \ref{sdulor}
following the Connes' guidelines we managed to obtain an
{\em evaluation} rather than the {\em expression} for the distance.
In 'good' cases, in particular, in the Minkowski spacetime, this
evaluation is exact and gives an analog of the Connes formula.

The attempt to generalize it to arbitrary Lorentzian manifolds
resulted in building of counterexamples which show the drastic
difference between the Riemannian and Lorentzian manifolds. In
section \ref{sdupr} the duality principle was introduced in order
to point out the class of Lorentzian manifolds being as 'good' as
Riemannian ones. It turned out that one can find a 'bad' spacetime
even among those conformally equivalent to the Minkowskian one. An
example is provided in section \ref{sdupr}.

The Connes duality principle play an important r\^ole in the
framework of the so-called 'spectral paradigm' in the account of
non-commutative differential geometry \cite{connes}. However, the
correspondence principle for this theory is corroborated on
Riemannian manifolds. Following these lines, in section \ref{salg}
we show a way to introduce non-comutative Lorentzian geometry.

\section{Connes formula}\label{scd}

Both the Riemannian distance and Lorentzian interval are based on
calculation of the same integral:

\[
\int_\gamma \sqrt{ds^2}
\]

\noindent which is always referred to a pair of points. The
intervals (resp., distances) as functions of two points are
obtained as extremal values of this integral over all appropriate
curves connecting the two points.

There is a remarkable duality to evaluate this integral suggested
by Connes for the Riemannian case. Consider it in more detail.

\medskip

For any two points $x,y$ of a Riemannian manifold $\mmm$ connected
by a smooth curve $\gamma$ the following evaluation of its length
$\ell (\gamma)$ takes place:

\be{eriem}
f(y) - f(x) \le \sup_{\mmm}{| \nabla f |}\cdot \ell (\gamma)
\ee

\noindent based on the Cauchy inequality:

\[
(\nabla f, \dot{\gamma}) \le |\nabla f| \cdot |\dot{\gamma}|
\]

So, the distance $\rho(x,y)$ between the points of the manifold
satisfies the following inequality

\[
\rho(x,y) \ge \sup (f(y) - f(x)),
\]

\noindent where $f$ ranges over all functions whose gradient does
not exceed 1: $| \nabla f | \le 1$.

\medskip

It was shown by Connes \cite{connesreal}, that as a matter of fact
no curves are needed to determine the distance, which may be
obtained directly as:

\be{parfeq}
\rho(x,y) = \sup_{| \nabla f | \le 1}(f(y) - f(x))
\ee

\medskip

The physical meaning of this result is the following: we can
evaluate the distance between the points measuring the difference
of potentials of a scalar field whose intensity is not too high.
So, the following {\em duality principle} takes place in Riemannian
geometry:

\[
\sup (f(y) - f(x)) = \inf \ell (\gamma)
\]

Note that this formula is valid even for non-connected spaces: in
this case both sides of the above equality are equal $+\infty$, if
we assume, as usually, the infinite value of the infimum when the
ranging set is void.

The question arises: can we write down a similar evaluation for the
Lorentzian case?

\section{Duality inequality in Lorentzian geometry}\label{sdulor}

The Cauchy inequality from which the Riemannian duality principle
was derived is no longer valid in the Lorentzian case. Instead, we
have the following:

\be{eanticauchy}
(\nabla f, \dot{\gamma})^2 \ge (\nabla f)^2 \cdot (\dot{\gamma})^2
\ee

\noindent where both $\nabla f$, $\dot{\gamma}$ are non-spacelike:

\[
(\nabla f)^2 \ge 0, \qquad (\dot{\gamma})^2 \ge 0
\]

\noindent If under these circumstances we also have $(\nabla f,
\dot{\gamma}) \ge 0$, the inequality (\ref{eanticauchy}) reduces to

\[
(\nabla f, \dot{\gamma}) \ge |\nabla f| \cdot |\dot{\gamma}|
\]

Now let $x,y$ be two points of a Lorentzian manifold $\mmm$ such
there is a causal curve $\gamma$ going from $x$ to $y$. Then for
any global time function $f$ on $\mmm$ we immediately obtain the
analog of the inequality (\ref{eriem})

\[
f(y) - f(x) \ge \inf_{\mmm}{| \nabla f |}\cdot \ell (\gamma)
\]

\medskip

Introduce the class $\fff(\mmm)$ of global time functions
satisfying the following condition:

\be{eclassf}
(\nabla f)^2 \ge 1
\ee

Then the Lorentzian interval between $x$ and $y$

\be{elab}
l(x,y) = \sup \int_{\gamma} \sqrt{ds^2}
\ee

\noindent can be evaluated as follows:

\[
f(y) - f(x) \ge l(\gamma)
\]

\noindent provided the class $\fff$ (\ref{eclassf}) is not empty.
Introducing the value

\[
L(x,y) = \inf_{f \in \fff} (f(y) - f(x))
\]

\noindent we obtain the following {\em duality inequality}:

\be{eduineq}
l(x,y) \le L(x,y)
\ee

It is worthy to mention that this inequality is still meaningful
when the points $x$, $y$ can not be connected by a causal curve. In
this case the supremum $l(x,y)$ is taken over the empty set of
curves and its value is, as usually, taken to be $-\infty$, that is
why the inequality (\ref{eduineq}) trivially holds.

\medskip

Let us thoroughly describe this construction in the case when
$\mmm$ is the Minkowskian spacetime. The following proposition
holds:

\paragraph{Proposition.} Let $\mmm$ be Minkowskian spacetime.  Then
$L(a,b) = l(a,b)$ for any pair of points $a,b \in \mmm$.

\medskip

\proof Assume with no loss of generality that $a=0$.  If $b$ is in
the future cone of $a=0$, then $l = (b,b)$ is realized on the
segment $[0,b]$.  The value of $L$ is achieved on the function
$f(x) = (b,x)/\sqrt{(b,b)}$. Let $b$ be a future-directed isotropic
vector, then $l = 0$. For any $\epsilon$ such that $0< \epsilon <
1$ consider the function

\[
f_\epsilon (x) =
\frac{((1-\epsilon)b + \epsilon v,
x)}{\sqrt{\epsilon(1-\epsilon)(b,v)}}
\]

\noindent where $v$ a vector defining the time orientation. The
direct calculation shows that $(\nabla f)^2 \ge 1$ and $(\nabla
f,v)\ge 0$, that is, $f_\epsilon \in \fff$. In the meantime $f(b) =
\sqrt{\epsilon(v,b)/(1-\epsilon)}$ which can be made arbitrarily
close to zero by appropriate choice of $\epsilon$.

Now let the point $b$ be beyond the future cone of the point $0$,
therefore they can be separated by a spacelike hyperplane
$f_k(x) = (k,x) = 0$ and choose the vector $k$ to
be future-directed. Then $f_k(b) < 0$. Since $f_{\lambda k} (b) =
\lambda f_k \in \mmm$, the infimum is $-\infty$. In the
meantime $l(0,b) = -\infty$ as well since there are no
future-directed non-spacelike curves connecting $0$ with $b$.

\eprf

\paragraph{Remark.} Note that if we borrow the definition of
$l(a,b)$ from \cite{bimerl}, namely $l(a,b)=0$ for $b\not\in J^+(a)$,
then the duality principle will not hold even for the Minkowskian
case: this was the reason for us to introduce the definition
(\ref{elab}).

\medskip

Consider one more example. Let $\mmm = S^1 \times \rrr^3$ be a
Minkowskian cylinder where $S^1$ is the time axis. In this case any
two points $x,y \in \mmm$ can be connected by an arbitrary long
timelike curve, therefore $l(x,y)=+\infty$. In the meantime the
class $\fff(\mmm)$ is empty (since there is no global time
functions), and therefore $L(x,y)=+\infty$. So we see that even in
this "pathological" case the duality principle is still valid.

\medskip

Note that the class $\fff(\mmm)$ itself characterizes spacetimes.
In general, if the class $\fff(\mmm)$ is not empty, the spacetime
$\mmm$ is chronological (it follows immediately from that
$\fff(\mmm)$ consists of global time functions).

Now we may inquire whether all Lorentzian manifolds are as 'good'
as Minkowskian? In the next section we show that the answer is no.

\section{Duality principle}\label{sdupr}

\begin{figure}[ht]
\unitlength1mm
\begin{center}
\begin{picture}(60,40)
\thicklines
\put(25,25){\line(1,-1){10}}
\thinlines
\put(0,20){\vector(1,0){60}}
\put(30,0){\vector(0,1){40}}
\put(30,10){\circle*{1}}
\put(31.5,10){\mbox{$a$}}
\put(30,30){\circle*{1}}
\put(31.5,30){\mbox{$b$}}
\put(25,25){\circle*{1}}
\put(35,15){\circle*{1}}
\end{picture}
\end{center}
\caption{An example where the duality principle is broken.}
\label{fig1}
\end{figure}
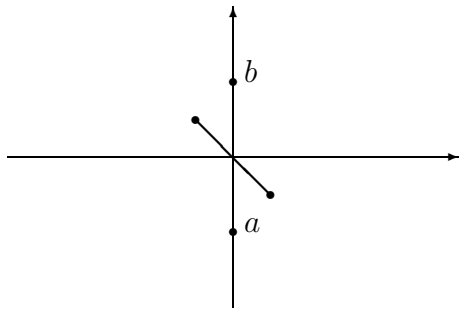

In the previous section we have proved the duality inequality
(\ref{eduineq}) which is always true in any Lorentzian manifold.
However, unlike the case of Riemannian spaces, this inequality may
be strict, which is corroborated by the following example. Let
$\mmm$ be a Minkowskian plane from which a closed segment
connecting the points $(1, -1)$ and $(-1, 1)$ is cut out (Fig.
\ref{fig1}).

Consider two points $a=(-2.0)$ and $b=(2,0)$. They can not be
linked by a timelike curve in $\mmm$, therefore $l(a,b) = -\infty$.
Meanwhile the class $\fff(\mmm)$ is not empty: it contains at least
the restrictions of all such functions defined on the whole
Minkowskian plane, thus $L(a,b)<+\infty$. Let us prove that the
value $L(a,b)$ is finite, supposing the opposite. If $L(a,b)$ would
be equal to $-\infty$, a function $f\in \fff(\mmm)$ should exist
for which $f(b)<f(a)$. Consider the behavior of the level line
$l_b$ of $f$ passing through the point $b$. Being spacelike, it can
not enter the cone $(-\infty < t < b; |x| \le |t-b|)$ which
contains the point $a$.  From the other hand, the point $a$ must
lie in the causal future cone $J_+(l_b)$, which is not the case.
So, $L(a,b) \neq l(a,b)$.

\medskip

Now, specifying the notion of 'good' spacetime, introduce the
duality principle.

\medskip

A Lorentzian manifold $\mmm$ is said to satisfy the {\em duality
principle} if for any its points $x,y$

\[
L(x,y) = l(x,y)
\]

\medskip

This characterization is global. The example we presented above
show that this notion is not hereditary: if we take an open subset
of a 'good' manifold it may happen that it will no longer enjoy the
duality principle.

Contemplating the above mentioned examples may lead us to an
erroneous conclusion that the reason for the duality principle to
be broken is when the spacetime manifold is not simply connected.
The next example \cite{krasnprivate} shows that there are manifolds
which are simply connected, geodesically convex, admit global
chronology but do not enjoy the duality principle.

Let $\mmm$ be a right semiplane $(-\infty < t < +\infty ; x > 0)$
with the metric tensor conformally equivalent to Minkowskian. It is
defined in coordinates $t,x$ as follows:

\[
g_{ik} = \frac{1}{x}
\left(
  \begin{array}{cc}
  1 & 0 \cr
  0 &-1
  \end{array}
\right)
\]

\medskip

The example illustrates the problems related with the dual
evaluations: it shows that the existence of a global time
function does not guarantee the class $\fff$ to be non-empty.
The spacetime $\mmm$ evidently admits global time
functions such as, for instance, $f(t,x) = t$.  However, the
following proposition can be proved.

\paragraph{Proposition.} The class $\fff(\mmm)$ is empty.

\medskip

\proof Suppose there is a function $f(x,t)$ satisfying
(\ref{eclassf}) and consider two values $A, B$ ($A<B$) of the
function $f$. The appropriate lines of constant level of $f$ are
the graphs of functions $t_A(x), t_B(x)$.  Since the derivative
$f_t>0$, we have $t_A(x)<t_B(x)$.  These functions are
differentiable and their derivatives are bounded:  $t_A', t_B' \le
1$ (because these lines are always spacelike), therefore they have
limits when $x \to 0$.  Let us show that these limits are equal.

Consider the difference $B-A$ and evaluate it:

\[
B-A = f(t_B(x),x) - f(t_A(x),x) =
\]
\[
 = \int_{t_A(x)}^{t_B(x)} f_t(t,x) dt \ge
\frac{1}{\sqrt{x}} \cdot (t_B(x) - t_A(x))
\]

\noindent where the first factor $1/\sqrt{x}$ is directly obtained
from the condition $(\nabla f)^2 \ge 1/x$.  So, the limit of
$t_B(x) - t_A(x)$ is to be equal 0. Since the values $A,B$ were
taken arbitrary, we conclude that all the lines of levels of the
global time function $f$ come together to a certain point.
Therefore these lines (being spacelike) cannot cover all the
manifold $\mmm$.

\eprf

This proposition shows that the space $\mmm$ does not support
duality principle: we can take two points $a,b$ on a timelike
geodesic and calculate the interval $l(a,b)$, while
$L(a,b)=+\infty$.

\section{ Algebraic aspects and quantization}\label{salg}

Let us study the dual evaluations from the algebraic point of view.
It was pointed out yet by Geroch \cite{geroch} that the geometrical
framework of general relativity can be reformulated in a purely
algebraic way. Recall the basic ingredients of Geroch's approach.

The starting object is the algebra $\aaa = \cinf(\mmm)$, then the
vector fields on $\mmm$ are the derivations of $\aaa$, that
is, the linear mappings $v: \aaa \to \aaa$ satisfying the Leibniz
rule:

\[
v(a \cdot b) = a\,v\cdot b + va\cdot b
\]

\noindent Denote by $\vvv$ the set of vector fields on $\mmm$
($=$ derivations of $\aaa$). It is possible to develop tensor
calculus along these lines: like in differential geometry, tensors
are appropriate polylinear forms on $\vvv$. In particular, the
metric tensor can be introduced in mere algebraic terms.

The Geroch's viewpoint is in a sense 'pointless' \cite{ps}: it
contains no points given {\em ab initio}. However the points are
immediately restored as one-dimensional representations of $\aaa$.
For any $x\in \mmm$ the appropriate representation $\hat{x}$ reads:

\[
\hat{x}(a) = a(x) \qquad a\in \aaa
\]

\medskip

Now let $\mmm$ be a Riemannian manifold. If we then decide to
calculate the distance between two representation $x,y$ in a
'traditional' way we have to introduce such a cumbersome object as
continuous curve in the space of representations. It is the result
of Connes (\ref{parfeq}) which lets us stay in the algebraic
environment:

\[
\rho(x,y) = \sup_{f\in \fff }(f(y) - f(x))
\]

\noindent and the problem now reduces to an algebraic description
of the class $\fff$ of suitable elements $f$ of the algebra
$\aaa$. The initial Connes' suggestion still refers to
points:  $\fff = \{ a\in \aaa |\, \forall m\in \mmm \: |\nabla
a(m)| \le 1 \}$.

Connes' intention was to build a quantized theory which could
incorporate non-commutative algebras as well. For that, the
construction of {\em spectral triple} was suggested \cite{connes}.

\medskip

A spectral triple $(\aaa, \hhh, D)$ is given by an involutive
algebra of operators $\aaa$ in a Hilbert space $\hhh$ and a
self-adjoint operator $D$ with compact resolvent in $\hhh$ such
that the commutator $[D,a]$ is bounded for any $a\in \aaa$ (note
that $D$ is not required to be an element of $\aaa$).

\medskip

Then for any pair $(x,y)$ of states ($=$ non-negative linear
functionals) on $\aaa$ the distance $d(x,y)$ between $x$ and $y$
may be introduced:

\[
d(x,y) = \{ |x(a) - y(a)| \,:\, a\in \fff \}
\]

\noindent with the following class of 'test elements' of the
algebra $\aaa$

\be{efnonc}
\fff = \{ a\in \aaa \:: ||[D,a]|| \le 1 \}
\ee

\medskip

The suggested construction satisfies the correspondence principle
with the Riemannian geometry. Namely, we form the spectral
triple with $\aaa= \cinf(\mmm)$, $\hhh = \lll^2(\mmm, S)$ --- the
Hilbert space of square integrable sections of the irreducible
spinor bundle over $\mmm$ and $D$ is the Dirac operator associated
with the Levi-Civit\`a connection on $\mmm$ \cite{semaz}. Then
$d(x,y)$ recovers the Riemannian distance on $\mmm$ (see, e.g.
\cite{connes}).

Comparing the definition (\ref{efnonc}) of the class $\fff$ with
that used in section~\ref{scd}: $\fff = \{ a\in \aaa |\, \forall
m\in \mmm \: |\nabla a(m)| \le 1 \}$ we see that the operator $D$
is a substitute of the gradient. Following \cite{gianni,froh}
the gradient condition (\ref{efnonc}) can be written in terms of
the Laplace operator taking into account that:

\[
(\nabla f)^2 =
\frac{1}{2}\Delta(f^2) - f\Delta f
\]

\noindent which restores the metric on $\mmm$ according to the
Connes' duality principle (\ref{parfeq}) for Riemannian manifolds.
However this condition is still checked at every point of $\mmm$.

We suggest an equivalent algebraic reformulation of (\ref{efnonc})
with no reference to points. Starting from the notion of the
spectrum of an element of algebra \cite{dix}:

\[
\mbox{\rm spec} (a) = \{\lambda \in \ccc |\, a-\lambda \cdot \1
\mbox{ is not invertible} \}
\]

\noindent and taking into account that the spectrum of the
multiplication operator coincides with the domain of the
multiplicator we reformulate the Connes' condition $f\in \fff$ as

\be{espectr}
\mbox{\rm spec} (\1 - (\nabla f)^2 ) \quad \mbox{ is non-negative}
\ee

Within this framework, to pass to Lorentzian case, we simply
substitute the Laplacian $\Delta$ by the D'Alembertian $\Box$, and
the spectral condition (\ref{espectr}) is changed to

\[
\mbox{\rm spec} ((\nabla f)^2 - \1) \quad \mbox{ is non-negative}
\]

\noindent which makes it possible to recover the Lorentzian
interval provided the duality principle holds.

\medskip

So we see that the notion of spectral triple is well applicable to
develop quantized Lorentzian geometry along the lines of Connes'
theory.

\section*{Acknowledgments}

The work was supported by the RFFI research grant (97-14.3-62). One
of us (R.R.Z.) acknowledges the financial support from the Soros
foundation (grant A97-996) and the research grant "Universities of Russia".

We would like to thank the participants of the research seminar of
the Friedmann Laboratory for theoretical physics (headed by
A.A.Grib), especially S.V.Krasnikov and R.Saibatalov, for profound
discussions.


\begin{thebibliography}{99}

\bibitem{bimerl} Beem, J., Erlich, P.,
Global Lorentzian geometry,
Marcel Dekker, New York,
1981

\bibitem{connes} Connes, A.,
Non-commutative geometry,
Academic Press, San Francisco,
1994

\bibitem{connesreal}
Connes, A.,
Non-commutative geometry and reality,
Journal of Mathematical Physics,
{\bf 36} (1995), 6194

\bibitem{dix}
Dixmier, J.,
Les $C^*$-alg\`ebres et leurs repr\'sentation,
Gauthier-Villars, Paris,
1964

\bibitem{froh} Fr\"ohlich, J., K.Gawedzki, A.Recknagel,
Supersymmetric quantum theory and (non-commutative) differential
geometry,
eprint hep-th/9612205

\bibitem{geroch} R.Geroch,
Einstein Algebras,
Communications in Mathematical Physics,
{\bf 26} (1972), 271

\bibitem{gianni} Landi, G.,
Introduction to noncommutative spaces and their geometry,
Springer-Verlag, Berlin,
1997

\bibitem{krasnprivate} Krasnikov, S.V.,
private communication

\bibitem{semaz} Palais, R.S.,
Seminar on the Atiyah-Singer index theorem,
Princeton, New Jersey,
1965

\bibitem{ps} Parfionov, G.N., Zapatrin, R.R.,
Pointless spaces in general relativity,
International Journal of Theoretical Physics,
{\bf 34} (1995), 737

\end{thebibliography}
\end{document}